\documentclass[a4paper,12pt]{article}
\usepackage{t1enc}
\usepackage[latin1]{inputenc}
\usepackage[english]{babel}
\usepackage{epsfig}
\usepackage{graphics}
\begin{document}
\centerline {\bf Implications of a solar-system population of massive  
4th generation neutrinos} 
\centerline{\bf   for underground searches of monochromatic 
neutrino-annihilation signals.} 
\vskip 0.2in
\centerline {\bf K.M. Belotsky$^1$, T. Damour$^2$, and  M.Yu. 
Khlopov$^{1,2}$}
\centerline {\it $^1$ Center for Cosmoparticle Physics "Cosmion",} 
\centerline {\it 125047, Moscow, Russia}
\centerline {\it  $^2$ Institut des Hautes Etudes Scientifiques,}
\centerline {\it 91440, Bures-sur-Yvette, France}
\vskip 0.4in


\centerline{\bf Abstract}

It has been recently pointed out that any primary galactic population of
Weakly Interacting Massive Particles (WIMP) generates, through collisions
with solar matter, a secondary  population of ``slow''
WIMPs trapped in the inner solar system. We show that taking into
account this ``slow'' solar-system population dramatically enhances
the possibility to probe the existence of stable massive neutrinos 
(of a 4th generation) in underground neutrino experiments. Though 
neutrinos, with mass in the 45-90 GeV range,
 can only represent a sparse subdominant 
component of galactic cold dark matter, a combination of enhancement factors
makes it possible to discriminate their contribution to WIMP 
annihilation effects in the Earth. Our work suggests that a reanalysis
of existing underground neutrino data should be able to bring 
extremely tight constraints on the possible existence of a
stable massive 4th neutrino. 

\section{Introduction}

The total number of quark-lepton families is not known for sure. 
Experimentally, three 
generations have been found at present. Theoretically, the number of 
generations could be larger.
In superstring-inspired particle models the number of generations is 
defined by some topological characteristics of the manifold of 
compactified additional dimensions, 
and any number of generations is a priori possible \cite{Green}. 

Another important prediction of 
almost all realistic superstring-inspired models is the existence of at 
least one additional 
$U(1)$ gauge group in the low energy limit of the theory. Recently 
\cite{Fargion99} it was 
suggested to ascribe such a new $U(1)$ gauge group to an additional, 
fourth, fermion generation 
only. In this case the new gauge group can remain unbroken. 
Such an unbroken $U(1)$ gauge group implies the existence of a strictly 
conserved charge, which, 
in turn, accounts for the stability of the lightest particle 
of the 4th generation, and forbids any mixing with the other (usual) 
three generations. It will 
be assumed that the lightest particle of the 4th generation is its 
neutrino.

The direct search for new generation fermions on accelerators leads to a
lower bound on their masses in the range 50-100 GeV. The mass of a 4th 
generation neutrino $m_\nu $ is restricted by the measurement of the 
width of the Z-boson to $m_\nu >m_Z/2\approx 45$ GeV if it is 
a (quasi)stable Dirac neutrino (for a Majorana one the restriction is 
slightly lower; for an unstable one the lower 
limit is about 90 GeV). Another possibility to search for a fourth 
generation neutrino, with 
 mass $m_\nu >m_Z/2 \approx 45$ GeV, at accelerators was suggested in 
\cite{Fargion99}, 
\cite{Fargion96}. A detailed analysis of the data on the parameters of the
Standard Model, accounting 
for the possible contributions of virtual new generation fermions, 
allows for the existence of 
an additional generation if the mass of the new (fourth) neutrino is 
about 50 GeV \cite{Maltoni} 
and if the masses of the other fermions of the new generation exceed 100 
GeV. 

The existence of new generation fermions in the Universe can lead to 
many observable astrophysical effects. This makes the appropriate 
cosmological and astrophysical analysis an important tool for probing 
the possible existence and properties of such 4th generation fermions. 
In particular, a new neutrino, being plausibly the lightest of its 
generation, and thus possibly stable, is of the most interest in such an 
analysis. 

As was found long ago \cite{Gershtein}, the existence of a heavy Dirac 
neutrino is compatible with the upper limit on the total density of the 
Universe if its mass exceeds 2 GeV. Indeed, for a neutrino mass greater than 3 MeV, 
neutrino annihilation through 
weak interactions reduces their cosmological concentration at freeze 
out. A larger neutrino mass 
corresponds to a larger annihilation cross section, and therefore to a 
smaller relic heavy 
neutrino density   $\Omega _\nu = \rho_\nu / \rho_{critical}$ (see Fig.2 
below). Moreover, if 
$m_\nu = m_Z/2$ the Z-boson annihilation resonance leads to a dip in the 
relic density. 
For instance, if $m_\nu =50$ GeV, close to the Z-boson resonance dip, 
$\Omega _\nu \approx 10^{-4}$. Such a rare population of neutrinos does 
not play any significant 
dynamical cosmological role as a Cold Dark Matter (CDM) contributor. 
Nevertheless a more refined 
astrophysical analysis \cite{Fargion99}, \cite{Golubkov} showed that the 
effects of such rare, 
massive neutrinos can be accessible to some experimental searches and/or 
astrophysical 
observations. In the present paper, we shall mainly consider the mass 
range $ 45 < m_\nu < 90$ GeV, corresponding to such a sparse population 
of heavy neutrinos.

Heavy neutrinos, as any form of CDM, must be concentrated in galaxies. 
The ratio 
of galactic neutrino density to the mean cosmological density is a 
model-dependent parameter 
(denoted here as $\xi $) which is strongly sensitive to the details of 
galactic halo formation. 
In this work we shall assume that the concentration factor $\xi $ is the
same for massive neutrinos and for the dominant contributor to CDM.
It is usual to estimate $\xi $ by taking the ratio between a 
local (i.e, in the vicinity of the Solar system) density of CDM equal to
$ 0.3 \,\rm{GeV/cm^{-3}} $, and a mean cosmological density corresponding
to $\Omega _{CDM}=0.3$. This leads to the ``standard'' estimate:
$\xi = 2 \times 10^{5}$. In most of this paper, we shall assume 
this standard value (except, when explicitly said otherwize).

In previous papers \cite{Fargion99}, \cite{Golubkov} we 
already analyzed several astrophysical consequences of such a sparse 
galactic population of 
heavy neutrinos. In particular we studied \cite{monochromatic} the 
(ordinary) neutrino fluxes 
emitted by the annihilation of heavy neutrino-antineutrino pairs 
accumulated in the core of 
the Earth. [Such an accumulation takes place for many types of weakly 
interacting massive 
particles (WIMP) \cite{wimp}.] 

Recently, it was pointed out \cite{Bergstrom} that 
this annihilation flux can be strongly enhanced by the existence of a 
``slow'' Solar-system 
population of WIMP's, trapped in the gravitational field of the Solar 
system by an initial 
inelastic interaction with the Sun \cite{Damour}. The aim of the present 
paper is to analyze
in detail, in the case of a heavy neutrino WIMP, the density of this new 
``slow'' Solar-system 
population, and its enhancement effect on the annihilation flux from the 
core of the Earth. We shall show that the existence of the slow population 
qualitatively improves the sensitivity of underground neutrino data to 
the effects of 4th neutrino annihilation, and makes these data a 
significant probe of the existence of a 4th neutrino.

\section{Number density of the ``slow'' heavy neutrino population}

Let us recall (from \cite{Damour}) that the ``slow'' population of WIMPs 
is generated by inelastic collisions of incident primary (``fast'')
galactic WIMPs with nuclei in the outer 
layers of the Sun. 
A fraction of these WIMPs are scattered, by the collision, on orbits 
that ``graze'' the surface 
of the Sun, and which evolve, under the subsequent perturbing 
gravitational influence of planets, 
onto orbits that do not penetrate the Sun. This allows these WIMPs to 
survive for a long time 
in the Solar system. This population has a much lower typical velocity 
($\sim $30 km/sec) than 
the incident galactic one. This fact amplifies the probability of their 
capture by the Earth 
(we will further refer to this population as ``the slow component''). 
The number density (in 
the vicinity of the Earth) of this slow component of heavy neutrinos, 
$n_{slow}$, has been 
derived in Ref.\cite{Damour} and reads:

\begin{eqnarray}
n_{slow}=\frac{0.212}{v_0/220\,\rm{km/s}}g_{tot}^{(-10)}\times n_{gal},\\ 
\nonumber
g_{tot}^{(-10)}=10^{10}\times \sum \limits_A \frac{f_A}{m_A}\sigma 
_AK_A^s\,(\rm{GeV^3}),\\ \nonumber
K_A^s=\frac{v_0}{\beta_+}\int_0^{v_{max}}4\pi v_\infty dv_\infty 
f(v_\infty)F^2(v_\infty).
\end{eqnarray}

Here $f(v_\infty)$ is the angular average of the velocity distribution of
galactic WIMPs. [Our notation differs slightly from Ref. \cite{Damour}
in that we factor out the number density $n_{gal}$ from the
phase-space distribution function, and delete the bar over $f(v_\infty)$
indicating the angular average.] The parameter $v_0$ is the 
characteristic velocity entering the (assumed) Maxwellian velocity 
distribution $f(v_\infty )$. Throughout the 
paper the index $\infty $ refers to quantities far from the Sun or 
(depending on the context) the Earth. The meaning of the other quantities 
entering the equations above is: $m_A$ is the mass of the nucleus $A$,  
$f_A$  denotes its mass fraction in the Sun, $\sigma _A$ is the 
cross-section of neutrino-nucleus scattering in the point-like 
approximation. The form-factor $F(v_\infty)$ takes into account the 
effect linked to the extension of the nucleus. More about cross-section 
and form-factor below. The quantity $g_{tot}$ (which has dimensions
$[cross section] \times [mass]^{-1}$ should be expressed in GeV$^3$. 
Here and below we use the notation

\begin{equation}
\beta _{\pm }=\frac{4m_\nu m_A}{(m_\nu \pm m_A)^2},
\end{equation}

$m_\nu $ denoting the neutrino mass. The upper limit,  $v_{\max }$, of 
the integral above comes 
from kinematics and is read from Eq.(2.16) of \cite{Damour}

\begin{equation}
v_{\max }=\frac{\sqrt{4m_\nu m_Av_{esc}^2(\bar{R}_S)-(m_\nu 
+m_A)^2\alpha }}
{\left| m_\nu -m_A\right| },
\end{equation}

in which we used the following numerical estimates: 

\begin{equation}
v_{esc}^{2}(\bar{R}_S=0.907R_S)=(644\,\rm{km/s})^2,
\end{equation}

where $R_S$ denotes the Solar radius, and $\alpha = G M_S/a  \simeq 
(30\,\rm{km/s})^2$, $a$ denoting the semi-major axis of the WIMP orbit.
 As said above, we take a 
WIMP velocity distribution (far from the Sun) which is Maxwellian, i.e.
(after angular averaging)

\begin{equation}
f(v)=\frac 1{4\sqrt{\pi ^3}v_0v_Sv}\left( \exp \left( 
-\frac{(v-v_S)^2}{v_0^2}\right) -
\exp \left( -\frac{(v+v_S)^2}{v_0^2}\right) \right). 
\end{equation}

Here $v_S$ denotes the velocity of the Sun relative to the gas of 
galactic neutrinos. In the 
estimates below, we took $v_S=200$ km/s.  We also estimated the effect 
of varying  $v_S$ between 
zero and  $200$ km/s and found that it changed only insignificantly the 
results given below. 
The parameter $v_0$ denotes the characteristic velocity dispersion of 
the Maxwell distribution 
(i.e. it measures the ``temperature'' of this distribution). In our 
estimates below we shall 
use the value $v_0 = 220$ km/s. For comparison, we shall also mention
 the results for 
$v_0 = 266$ km/s, which corresponds to a mean velocity of 300 km/s.

The data about the chemical composition of Solar matter was taken as in Ref. 
\cite{Damour} (i.e., a combination of the values given in 
\cite{Jungman} and \cite{Bahcall}). We also follow Ref. \cite{Damour}
in taking nuclei form-factor of the form:

\begin{equation}
F^2(v)= \exp\left(-\frac{Q(v)}{Q_A}\right), 
\end{equation}

where $Q=\frac{m_\nu (v^2+\alpha )}2$ is the transferred energy from 
neutrino to nucleus,
 $Q_A=\frac{3\hbar ^2}{2m_AR_A^2}$, 
$R_A=10^{-13}\,{\rm cm}\,(0.3+0.91(\frac{m_A}{\rm{GeV}})^{1/3})$.

The total $\nu -A$ cross-section for a point-like nucleus $\sigma _A$ is 
approximated by a 
purely spin independent coupling, except in the case of hydrogen. Since 
most contributing nuclei 
have a first excited energetic level higher than the characteristic 
energy tranfer $Q$ 
the collision is assumed to be elastic. This leads to a cross-section of 
the form

\begin{equation}
\sigma _A=\frac{G_F^2}{8\pi }\frac{m_\nu ^2m_A^2}{(m_\nu 
+m_A)^2}\left(F_V^2+3F_A^2
\delta _A^1\right), 
\end{equation}

where $G_F$ is the Fermi constant, $F_V=-(A-2(1-2\sin ^2\theta _W)Z)$ is 
the vector form-factor (all $q^2$-dependence is taken out into $F(v)$), 
$\theta _W$ is the Weinberg angle, and $F_A$ is the axial form-factor, 
which is taken into account only in the case of  hydrogen (hence the 
Kronecker symbol  $\delta _A^1$).

After substituting all these quantities in Eq. (1) we obtain the 
dependence on $m_\nu $ of the density enhancement
ratio $\delta_E = \frac{n_{slow}}{n_{gal}}$ showed in Fig.1.
In Table 1, we fix $m_{\nu} = 50$ GeV, and study the sensitivity
of the density enhancement $\delta_E$ to the choices of the two
velocity parameters entering our study: $v_0$ and $v_S$.

\begin{figure}
\begin{center}
\centerline{\epsfig{file=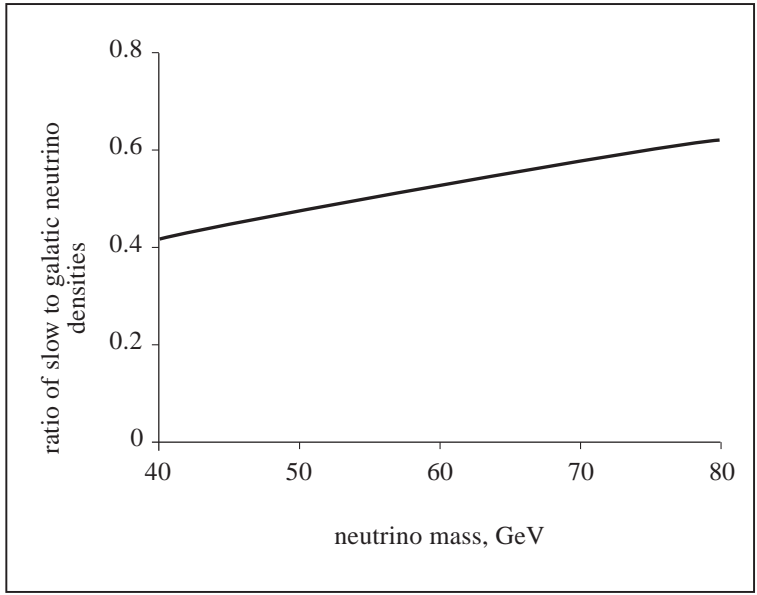,width=10cm}}
\caption{Density enhancement $\delta_E = n_{slow}/n_{gal}$
 as a function of the neutrino mass $m_\nu$.}
\end{center}
\end{figure}

\begin{table}[h]
\begin{tabular}[]{|c|c|c|}
\hline & $v_0=220\,\rm{km/s}$ &  $v_0=266\,\rm{km/s}$ \\ \hline 
$v_S=0$ &  0.728  & 0.529 \\ \hline
$v_S=200\,\rm{km/s}$ & 0.475 & 0.388 \\ \hline
\end{tabular}
\caption{Density enhancement $\delta_E = n_{slow}/n_{gal}$
 for different velocity parameters.}
\end{table}

The crucial new result to note from Fig. 1 and Table 1 is that, contrary 
to the case of a generic 
neutralino WIMP where the density enhancement (near the Earth) due to 
the slow population, 
$\delta_E = n_{slow}/n_{gal}$, was typically of the order of a few 
percent, here this density 
enhancement is of the order of  50 \%. This shows that it is crucial to 
take into account the 
existence of the slow population to estimate the detectability of a 4th 
neutrino . In addition 
to this significant enhancement in the local density of WIMPs, we turn 
now to the estimate of 
the capture rate of the WIMPs by the Earth which is further amplified by 
the fact that slow WIMPs 
(with velocity $\sim 30$ km/s) are much more easily captured by the 
Earth gravitational field 
than galactic ones (with velocity $\sim 220$ km/s).

\section{Capture rate of massive neutrinos by the Earth}

The neutrinos interacting with the Earth matter can be captured by the 
Earth 
gravitational  potential well if they lose enough energy in the 
collision with a nucleus in 
the Earth.  The ``slow'' neutrinos, having an incident velocity already 
comparable with the escape 
velocity from the Earth ( $v^{esc}_E \sim 10\,\rm{km/s}$), have a much 
greater probability to be captured 
than galactic ones. It is therefore quite important to take their effect 
into account. 

Among all nuclei present in the Earth the iron nuclei play the main role 
in the capture process for the 
neutrino mass range ( $ 45 \,\rm{GeV} < m_{\nu} < 90 \,\rm{GeV}$) that 
interests us.

The capture rate of WIMPs by the Earth has been studied in detail by 
Gould \cite{Gould} (see also \cite{Bergstrom}). In the present study,
 we estimated the capture rate by starting from Eq.(5.7) and Eq.(5.8) of
\cite{Bergstrom}, and by approximating the result in the following way.
We replaced the integrals over the volume of the Earth of the type
$\int d^3x\,n_A(r) f(r)$ by $N_A \langle f(r)\rangle$, where $N_A$
is the total number of nuclei of type $A$, and where  $\langle f(r)\rangle$
is a suitable (approximate) average of $f(r)$. Actually, we consider
only the capture by iron. As iron is concentrated in the core of the
Earth we can use the average of the squared escape velocity given 
in \cite{Gould}: $\langle v_{esc}^2 \rangle=1.6\times
(11.2\,\rm{km/s})^2$. We assume that the fraction of the mass of the Earth
in iron is 20 \%. Finally, our simplified capture rate reads:

\begin{eqnarray}
R_{cap}=\frac{2Q_{Fe}}{\beta_-m_{\nu}}N_{Fe}\sigma_{Fe}\int_0^{\sqrt{
\beta_-\langle v_{esc}^2 \rangle}} \frac{dn_{\nu}}{v}\times \nonumber \\
\times\left[ \exp\left(-\frac{m_{\nu}}{2Q_{Fe}}v^2\right)
-\exp\left(-\frac{m_{\nu}}{2Q_{Fe}}\beta_+(\langle v_{esc}^2 
\rangle+v^2)\right)\right]
\end{eqnarray}

The neutrino distribution function $dn_{\nu}$ to be inserted in Eq.(8) 
depends on whether we consider galactic or slow neutrinos.
 For galactic neutrinos it is
$dn_{gal}=n_{gal} \, f(v)d^3v$. For slow neutrinos we followed
Eq.(3.15) of \cite{Bergstrom} (with $\lambda =1$ ; $\varepsilon =0.18377$), 
i.e, explicitly,

\begin{eqnarray}
dn_{slow} = n_{slow}\, f_{slow}(v_{slow})dv_{slow} 
 =  n_{slow}\, Cv\frac{\theta (1.617\times v_E-v)}{(3\times 
v_E^2-v^2)^{0.6}}\times \nonumber \\
\times \left( \sqrt{v^2-0.816\times v_E^2}\,
\theta (v-0.904\times v_E) 
 -\sqrt{v^2-1.184\times v_E^2}\,\theta (v-1.089\times v_E) \right) dv \, ,
\end{eqnarray}

where $C=91.775^{-1}$ is a normalization factor, $\theta (x)$ is the 
step function, $v_E=30\,$km/s is the velocity of the Earth orbital 
motion, and where $n_{slow}$ is given by Eq.(1) above. 

Another crucial parameter for our problem is the value 
of the incident galactic neutrino number density itself. Contrary to the case 
usually considered for WIMP 
capture, we cannot assume here that the sparse 4th neutrino population 
is the principal constituent 
of Cold Dark Matter, i.e we cannot assume that its galactic density has 
the ``standard'' value 
$\rho_{CDM} \sim 0.3\,\rm{GeV/cm^3}$. Instead, we need to estimate what is 
the relic density of 4th 
neutrinos left over from the Big Bang, and to multiply it by an estimate 
of the typical density enhancement,
 in a galaxy. We can then write the galactic neutrino 
number density (as a function of 
the neutrino mass) as:

\begin{equation}
n(m_\nu )=\xi \: n_{rel}(m_\nu ).
\end{equation}

Here $\xi $ denotes the parameter of neutrino clustering in the Galaxy, 
i.e. the ratio of the 
local neutrino density to the relic one, 
$n_{rel}(m_\nu )$. As already mentioned above, in our estimates
we take the standard value $\xi = 2\times 10^5$.

As for the relic number density (of neutrinos and antineutrinos), we 
take the estimate 
derived in \cite{Fargion99}, namely:

\begin{equation}
n_{rel}(m_\nu )=\frac{4.2\times 10^{-18}}{\sqrt{g_{*}}m_pm_\nu 
\left\langle
\sigma _{ann}\beta \right\rangle }\left( 40+\ln \left( 
\frac{g_s}{\sqrt{g_{*}}}m_pm_\nu 
\left\langle \sigma _{ann}\beta \right\rangle \right) -\ln \sqrt{\frac{m_\nu }{T_*}}
\,\right) \times n_\gamma \,.
\end{equation}

Here $g_{*}\approx 80$ is the number of effective degrees of freedom at the 
moment of  neutrino 
freeze-out, $g_s=2$ is the number of neutrino spin states, $n_\gamma 
=411\,\rm{cm^{-3}}$ is 
the present number density of relic photons, $\left\langle \sigma 
_{ann}\beta \right\rangle $ 
is the thermally averaged annihilation cross-section (multiplied by the 
dimensionless neutrino 
relative velocity) at freeze out. The value of the last logarithm entering 
the above result can be 
taken to be 1.6 in the mass range $45 \,\rm{GeV}< m_\nu <90$ GeV  that 
we shall consider. In this mass 
range, the dominant annihilation channel for neutrinos is the channel 
involving one intermediate 
$Z$-boson. In this case the value $\left\langle \sigma_{ann}\beta 
\right\rangle $ may be taken 
(for the  non-relativistic case) in the form

\begin{equation}
\left\langle \sigma _{ann}\beta \right\rangle =\sigma _{ann}\beta 
^{*}\sqrt{2}=
Br^{-1}(Z\rightarrow \nu _e\nu _e)\frac{\bar g^4}{2^8\pi }\frac{m_\nu^2}
{(4m_\nu ^2-m_Z^2)^2+m_Z^2\Gamma _Z^2}\sqrt{2},
\end{equation}

where $Br(Z\rightarrow \nu _e\nu _e)\approx 0.0667$ is the branching 
ratio of the decay of 
the Z-boson into two electron neutrinos, $m_Z$ and $\Gamma _Z$ are the 
mass and width of 
the Z-boson, and $\bar g=\sqrt{4\sqrt{2}G_Fm_Z^2}$ is the dimensionless 
constant of weak interaction. 
Note that, if we were to consider larger neutrino masses ($m_\nu > 90$ 
GeV), the channel 
$\nu \tilde \nu \rightarrow W^{+}W^{-}$ would start to dominate and the 
cross-section would 
increase with $m_\nu$.

The relic neutrino density in units of the critical density is shown in 
Fig.2. Note in particular that, when $m_\nu =50$ GeV, $\Omega _\nu 
=1.7\times 10^{-4}$ 
(which is indeed much smaller than  the usually considered WIMP relic 
densities). Still in 
this numerical example of $m_\nu =50$ GeV, by multiplying this relic 
density by the 
amplification factor $\xi = 2 \times 10^5$ mentioned above, we find a 
local, galactic number 
density of massive neutrinos equal to about $0.2\,\rm{MeV/cm^3}$. This is 
indeed much smaller than 
the ``standard'' value for usual WIMPs.

\begin{figure}
\center{\epsfig{file=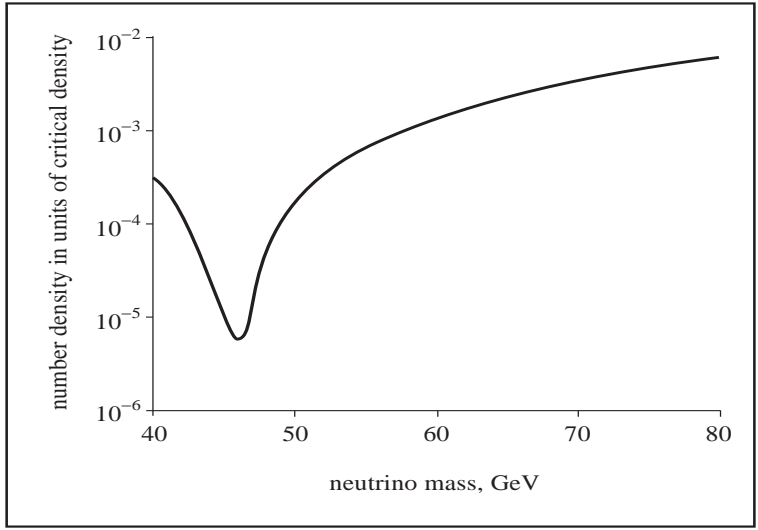,width=10cm,height=7cm}}
\caption{The density of primordial neutrinos in  units of the 
critical density.}
\end{figure}

After their capture, neutrinos (and antineutrinos) are thermalized, 
settling in the Earth core, 
and neutrino-antineutrino pairs start to annihilate. If the equilibrium 
between capture and 
annihilation is reached, the annihilation rate is equal to the capture rate. 
Let us estimate the 
capture rate ($R_{eq}$) for which the equilibrium between capture and 
annihilation inside the 
Earth is reached during the lifetime $t_E$ of the Earth. The variation 
of the number of 
accumulated neutrinos and antineutrinos $N$ satisfies

\begin{equation}
\frac{dN}{dt}=R_{cap}-R_{ann},
\end{equation}

where  $R_{cap}$ and $R_{ann}$ denote the capture and annihilation rates 
respectively. We have

\begin{equation}
R_{ann}=2\int \frac{n^2}4\left\langle \sigma_{ann}v_{rel}\right\rangle dV.
\end{equation} 

Here the  factor 2 accounts for the fact that in one annihilation a pair 
of neutrinos disappears; 
the factor $1/4$  comes from the fact that $n$ denotes the total 
(neutrinos plus antineutrinos) 
neutrino density within the thermalized core of the Earth; in 
$ \left\langle \sigma _{ann}v_{rel}\right\rangle $, $v_{rel}$  denotes 
the relative velocity of 
thermalized neutrinos. A rough estimate of $R_{ann}$ is obtained by 
assuming a homogeneous 
distribution of thermalized neutrinos within a certain 
volume $V_{therm}$. We define  $V_{therm}$ as 
the volume bounded by the radius which can be reached by a particle of 
kinetic energy $T_{therm}$ 
in the center of the Earth freely moving in the Earth potential well. 
The typical neutrino energy $T_{therm}$ 
is supposed to correspond to the 
 temperature of the Earth core. The latter is not known exactly, but
 is around 10000 K. We can therefore consider  $T_{therm}=1$ eV
 as fiducial value. This yields:

\begin{equation}
V_{therm}=\frac 43\pi 
R_{therm}^3,\,\,R_{therm}=\sqrt{\frac{2T_{therm}}{T_{esc0}}\frac
{\langle \rho_E\rangle }{\rho _{E\,core}}}R_E,
\end{equation}

where $T_{esc0}\equiv \frac{m_\nu v_{esc0}^2}2$, $\langle \rho_E\rangle 
=5.5\,\rm{g/cm^3}$ 
and $\rho _{E\,core} \approx 12 \,\rm{g/cm^3}$ are the mean and core Earth 
densities, and $R_E$ is 
the Earth radius. In this approximation we have

\begin{equation}
R_{ann}=\frac{N^2}{2V_{therm}}\left\langle 
\sigma_{ann}v_{rel}\right\rangle \equiv CN^2.
\end{equation}

It is then easily checked that, neglecting the (slow, linear) variation 
with time of  
$R_{cap}$, the solution for the time variation of $N$, or equivalently  
$R_{ann}$, is 
$R_{ann}=R_{cap}\tanh ^2\left( \sqrt{CR_{cap}}\,t\right) $. The 
``equilibrium'' rate $R_{eq}$, 
for which a balance between capture and annihilation within the Earth 
lifetime $t_E$ is maintained,
is therefore estimated to be $R_{eq}=1/(C \, t_E^2)$. 

Substituting all the factors, we finally obtain 

\begin{equation}
R_{eq}=0.46\times 10^{13}\,{\rm s^{-1}}\,\frac{(4m_\nu ^2-m_Z^2)^2+m_Z^2\Gamma 
_Z^2}{m_\nu ^{7/2}}\left(
\frac{T_{therm}}{1\,\rm{eV}}\right) ^{3/2}\left( \frac{\langle \rho _E\rangle }
{\rho_{E\,core}}\right) ^{3/2}.
\end{equation}

Here the mass of the neutrino, as well as  the mass and the width of the 
$Z$-boson are in GeV.
 For numerical estimates, we took $\frac{\rho _{E\,core}}{\langle 
\rho_E\rangle }=2$, and (as said above) 
$T_{therm}=1$ eV.

Having in hands, ready for comparison, the ``equilibrium'' rate $R_{eq}$,
we now come back to the estimate of the capture rate, as given by Eqs.(8)-(11)
above. Note that the capture rate of both galactic and slow neutrinos 
are proportional to the assumed value of the concentration factor $\xi$.


We plot in Fig.3
the total capture rate of neutrinos (summed over slow and galactic ones),
as a function of the neutrino mass, for $\xi=2\times10^5$. We also indicate what 
would be the capture rate if one included only galactic neutrinos.
 
 
 

\begin{figure}
\center{\epsfig{file=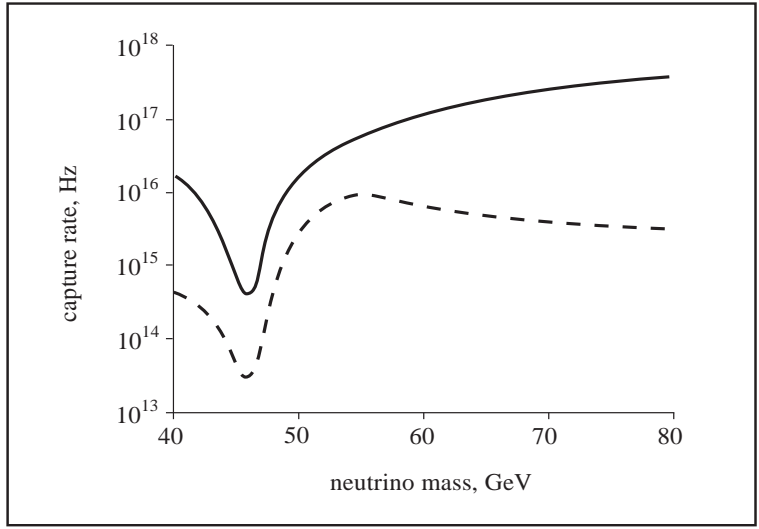,width=10cm,height=7cm}}
\caption{ Total capture rate of slow and galactic neutrinos (solid line),
compared to the capture rate of galactic neutrinos only (dashed line).}
\end{figure}

\begin{figure}
\center{\epsfig{file=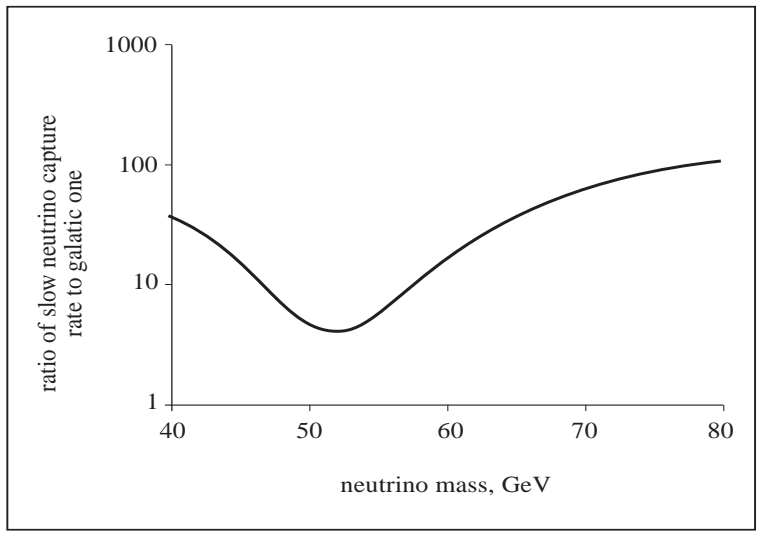,width=10cm,height=7cm}}
\caption{Ratio of the slow neutrino capture rate to galactic one.}
\end{figure}

The dip at $m_\nu \approx 45$ GeV reflects the resonant annihilation dip 
in primordial neutrino 
density (see Fig.2). The peak in the galactic neutrino capture rate at 
$m_\nu \approx 55$ GeV 
comes from the well known fact that when the WIMP mass is equal to the 
mass of iron nuclei their 
collision is efficient in slowing down the WIMPs. [This peak is shifted 
from the iron mass 
$m_{Fe} = 52$ GeV because of the rising relic neutrino density factor.] 
In the case of the slow 
WIMPs (which lose their kinetic energy during collisions much more 
effciently than the fast 
galactic WIMPs) this ``iron resonance'' effect is spread over a large 
range of masses and does 
not show up as a peak.

In Fig.4 we present the ratio of slow neutrino capture rate to the galactic one.
This ratio, obviously, is 
independent of galactic neutrino density. The dip around 
$m_{\nu} = m_{Fe} = 52$ GeV in Fig.4 is due to the "iron 
resonance" peak of the galactic neutrino capture. 

An important conclusion from Fig.3 is that, in the considered mass 
range, the equilibrium between 
capture ($R_{cap}$) and 
annihilation rates ($R_{ann}$) inside the Earth is established on a time 
scale smaller than the 
Earth lifetime. Therefore the annihilation rate can be simply taken as 
being equal to the (above 
computed) capture rate: $R_{ann}=R_{cap}$. 


To complete the results indicated in Fig.3, we give in Table 2 the 
galactic and slow neutrino 
capture rates  (in units of $10^{15}\,\rm{s^{-1}}$) for different velocity 
parameters, and for 
the special value $m_\nu =50$ GeV (and for the standard value 
$\xi = 2 \times 10^5$). Note that the equilibrium capture 
rate for 
$m_\nu =50$ GeV  amounts to $0.53\times 10^{13}\,\rm{s^{-1}}$.

\begin{table}[h]
\begin{tabular}[]{|c|c|c|c|c|}
\hline & \multicolumn{2}{c|}{$v_0=220\, \rm{km/s}$} & 
\multicolumn{2}{c|}{$v_0=266 \,\rm{km/s}$} \\
\cline{2-5} & galactic & slow & galactic  & slow \\
\hline $v_S=0$ & 3.86 & 18.6 & 2.88 & 13.50 \\
\hline $v_S=200 \,\rm{km/s}$ & 2.62 & 12.1 & 2.14 & 9.88 \\
\hline 
\end{tabular}\caption{Capture rates in units of $10^{15}\, \rm{s^{-1}}$ of  
galactic and slow neutrinos when 
$m_{\nu}=50$ GeV, for different velocity parameters.}
\end{table} 

One can see that our results are robust under varying the
 velocity parameters (retaining the 
Maxwellian form of the velocity distribution).

In the estimates above, we have not taken into account the effect of the 
so-called ``Sakharov 
enhancement'' of the annihilation rate due to the Coulomb-like 
attraction mediated by the new $U(1)$ 
interaction ascribed to neutrinos and antineutrinos of the 4th generation 
\cite{Sakharov}. This 
interaction has two effects: on one hand it reduces the relic neutrino 
density 
(by about 10\% if $\alpha _y=1/137$, and 25\% if $\alpha_y=1/50$), but 
on the other hand it strongly
increases the annihilation rate in the Earth. The overall effect is to 
strongly reduce the 
equilibration time scale between capture and annihilation in the Earth. 
Finally, if we were to take into account the
"Sakharov enhancement" we would be in the conditions of equilibrium for the 
whole neutrino mass range considered and for all the acceptable 
magnitudes of parameter $\xi $.

%

\section{Conclusions}

In the present paper we studied the capture by the Earth, and 
the annihilation in the Earth core, of 
hypothetical fourth neutrinos (candidate to a sparse sub-dominant 
component of galactic CDM). We took into account not only the primary
``fast'' population of neutrinos, but also the recently pointed out
secondary ``slow'' population \cite{Damour}.  It was found that the 
account of the  slow component is crucially important in the considered 
neutrino mass range, $ 45 < m_\nu < 90$ GeV. Indeed, the contribution  to
capture of the slow component is larger  by up to the two orders of magnitude 
than the one of the galactic component (Fig.4). 

These results suggest the 
crucial significance of underground experiments (AMANDA, 
Super-Kamiokande, Baksan and others) for testing the fourth neutrino 
hypothesis. 
For example, Ref. \cite{AMANDA} has derived the constraint
on WIMP annihilation in the Earth, from AMANDA data,
under some assumptions on the annihilation channels, and for a WIMP mass 
larger than 100 GeV. 
Making a rough extrapolation of  
this constraint down to a mass of about 50 GeV one finds a 
potential sensitivity of the already existing underground 
neutrino data to the annihilation of a 4th neutrino in the Earth,
for almost all the considered interval of neutrino mass. 
Note, that the presence of the slow component of a 4th neutrino plays 
a crucial role in 
this potential sensitivity. Of course, this example can 
serve only as an illustration, since special analysis of the
data in the framework of the hypothesis of a 4th neutrino is needed.

The possibility to distinguish, in underground neutrino 
experiments,  the contribution to annihilation effects of a sparse 
component of 4th neutrino from the contribution of 
other WIMPs (presumably dominating in the 
galactic CDM), results from the combination of several factors.
 The neutrino capture in the Earth is facilitated by the
relatively large weak interaction cross section of a massive neutrino and 
by the kinematic enhancement of neutrino momentum losses in 
collisions with iron nuclei. The neutrino annihilation effects in the Earth
are strengthened by the relatively large neutrino weak annihilation cross 
section near $Z$-boson 
resonance (which is further strongly enhanced by the Coulomb like effect
 of the new long range interaction), and by the existence of a monochromatic 
neutrino-antineutrino annihilation channel, specific to a 4th neutrino.

The presence of a slow component qualitatively enhances these factors. 
The slow component increases by up to 50\% the number density of 4th neutrinos 
near the Earth. Owing to their order-of-magnitude smaller mean velocity, 
the slow neutrinos are more effectively captured by the Earth (by up to
 two orders of magnitude) than the galactic ones. In the slow 
component capture the kinematic peak of iron nuclei capture is spread 
over the whole considered neutrino mass interval, making it accessible 
to experimental test. The establishment of kinetic equilibrium between 
neutrino capture and annihilation in Earth makes the predicted 
annihilation fluxes insensitive to the details of captured neutrino 
distribution. As a result, the account of the slow neutrino component 
makes the hypothesis of a stable massive 4th neutrino accessible to 
underground neutrino experimental tests even under the most unfavourable 
astrophysical conditions.

To conclude, our work shows that it is important to analyze existing
 (and future) underground neutrino data with the view of probing
 the hypothetical existence of a stable fourth generation 
neutrino with a mass about 50 GeV.  
The analysis of the data of MACRO, AMANDA, Kamiokande, Baksan and/or 
Super-Kamiokande is expected to provide an important probe of
(and probably stringent constraints on) this hypothesis, especially in the
case where one wishes to explain the DAMA event rate by assuming heavy 
neutrinos. The existence of a slow  4th neutrino component is crucial
in such an analysis because it strongly enhances the underground 
neutrino flux expected 
from 4th neutrino annihilation in Earth.
 This analysis can be viewed as a modest step towards 
the study of heterotic string phenomenology which generically leads
to the prediction of an additional $U(1)$, which, in turn,
 provides a motivation
for considering a {\it stable} 4th generation neutrino.

\section{Acknowledgement}

The work was performed in the framework of the project "Cosmoparticle 
physics" and was partially 
supported by the Cosmion-ETHZ and AMS-Epcos collaborations and by  a 
support grant for 
the Khalatnikov Scientific School.One of us (M.Yu.Kh.) expresses his 
gratitude to IHES for 
its kind hospitality and to ICTP for kind hospitality and help in the
finishing of this work.

\end{document}